\documentclass[a4paper,12pt,onecolumn,draftclsnofoot]{IEEEtran}
\usepackage{amssymb}
\usepackage{tabularx}
\usepackage{graphicx}
\usepackage{caption}
\usepackage{subcaption}
\usepackage[tbtags]{amsmath}
\usepackage{amsfonts}
\usepackage{mathrsfs}
\usepackage{multirow}
\usepackage[english]{babel}
\usepackage{cite}
\usepackage{url}
\usepackage{hyperref}
\hypersetup{
  colorlinks=true,
  linkcolor=red,
  linktoc=all,
  citecolor=blue,
  bookmarksnumbered=true
}
\usepackage{cleveref}
\usepackage{units}
\usepackage[section]{placeins}

\usepackage[colorinlistoftodos]{todonotes}

\IEEEoverridecommandlockouts

\begin{document}
\title{Potential 3rd COVID Wave in Mumbai: Scenario Analysis}
\author{
\IEEEauthorblockN{
TIFR Covid-19 City-Scale Simulation Team \\
Sandeep Juneja,
Daksh Mittal\footnote{Daksh Mittal is supported through the IDFC Institute} \\
}
{June, 2021} \\
}
\maketitle

\vspace*{-2.5cm}

\section{Summary}
The second wave of Covid-19 that started in mid-February 2021 in India is now subsiding with daily cases down from peak of 4,14,280 on May 6, 2021 to 48,768 on June 25, 2021. Similarly fatalities have also reduced from peak of 4529 on 10 May to 1183 on June 25, 2021. 
The second wave first arose in the state of Maharashtra,  and within it, Mumbai was amongst the first cities to see the
increase. Mumbai saw a peak case load of 11,163 on April 4, 2021 and peak fatalities of 90 on May 1, 2021. These numbers have come down to 648 on June 26, 2021 for the cases and 15 on June 26, 2021 for the deaths. See our analysis on the second Covid wave in Mumbai in \cite{May_report_2021}.

Increasingly the focus amongst the policy makers and general public is on the potential third wave.  When would it occur? Would it be as large or larger than the second wave? Would vaccines play a key role in ameliorating it? Would new, more infectious and virulent variants show up? Would they break through the immunity provided by the vaccines 
and by prior infection? Would reinfections increase in the coming months and contribute to this wave? Would laxity in population behaviour be the key driver for the third wave?

Many of these questions are difficult to predict through a model. How does one model the probability distribution of occurrence of the next variant, its infectiousness, and its virulence? This point was tragically illustrated in the second wave where the delta variant was the key driver, but this would have been near impossible to predict   late last year. Similarly, reinfections, especially severe ones, appear rare, but based on our current understanding, it is hard to accurately predict their future pattern.  Vaccines appear to be effective, but there may well be substantial breakthrough infections especially if new variants become significant. 

Due to these uncertainties, instead of projecting our {\em best guess scenario}, which we do not have, we conduct an extensive scenario analysis and track peak fatalities in the coming months in each of these scenarios \footnote{A draft version of this report was released in the last week of  June 2021, see \url{
http://www.tcs.tifr.res.in/~sandeepj/avail_papers/3rd\%20wave\%20SJ.pdf
}}.
 Peak fatalities can be compared with the past peaks and
 can  provide a reasonable guide to the
medical resources needed to successfully tackle the associated wave. 
Our other aim of  scenario analysis is to help policy-makers become  better aware of the factors that may drive
the potential third wave,  so that it can be caught and controlled  early.

While we discuss some of the  details later, our key conclusions are
\begin{itemize}
\item As per our model, about 80\%  of Mumbai population has been exposed to Covid-19  
by June 1, 2021 (see Figure~\ref{prevalence_affected_1}). Under the assumption  that all who are exposed have immunity against
further infection, it is unlikely that Mumbai would see a large third wave.
It is the reinfections that may lead to a large wave (see Figure \ref{peak_fatalities}). Reinfections could
occur because of declining antibodies amongst the infected
as well as by variants that can breakthrough the immunity provided by
prior infections. Mechanisms need to be in place that
continously measure emergence of reinfections and variants 
that can breakthrough existing immunity, including immunity provided through vaccines. 
\item
We also consider a somewhat pessimistic scenario
where 
\begin{itemize}
\item
reinfections are significant (amongst that 80\% recovered, 10\% 
are amenable to infections, and if infected,
 will follow the same disease progression as the first time infected),
 \item
 a new variant that is 50\% more infectious 
  and 50\% more virulent than the delta variant
  (we assume that delta variant is 2.25 times more infectious 
  than the earlier  variant in Mumbai and marginally more virulent),
  \item
   the vaccine effectiveness is only 30\%. 
\item
 the city is opened up  at 60\% level 
 and the  compliance is 20\% in non-slums and 10\% from June 2021 onwards.
\end{itemize}

 The resulting peak is seen to be 
  no larger then that under the second wave (see Figure \ref{peak_fatalities}, scenario
  VE =0.30, R=10\%, dark blue bar).
This may be true for other urban areas as well
where a large proportion of population has already been infected.
Regions that have more susceptible population
may witness larger waves when faced with a similar scenario.

Interestingly, the fact that a small increase in  susceptible population
 in presence of highly infectious variant leads
 to a relatively high increase in peak fatalities
 was illustrated through Figure 17 in \cite{May_report_2021}  (where reinfections lead to a small percentage increase
 in susceptible population but a relatively large increase in peak fatalities).
 This may  largely explain why 
 Mumbai saw a peak of around 11,000 cases a day in the second wave
 while Delhi saw a peak of around 28,000 and Bangalore of around 25,000.
 Mumbai,
 with a population of around
 13 million, already had  around 65\% population exposed to the disease (as per our model) on Feb 1.
 Delhi with 19 million population had an estimated 55\% population exposed
 around this time \cite{delhi_sero}. Bangalore, with a population
 similar to Mumbai had estimated 45\% exposed around Feb. 1
 (see \cite{bengaluru_sero}).  Thus, while the Mumbai administration
was well organised in tackling the second wave, the presence of high sero-positivity
may have  helped Mumbai avoid the tragic consequences that may have resulted from  
a much higher  peak.
 
 \item
 We further observe that under the scenario where the reinfections are mild so that they affect the fatality figures negligibly,
 where the new variants (beyond the existing delta variant) have a mild impact,
 it is natural for the city to further open up over time and for the population to become more lax over time.
 In this scenario we 
  observe a small wave under which  the fatalities begin to increase by around mid-August. See Figure~\ref{daily_deaths_1}. Hence, since the fatalities lag reported cases by around three weeks, the cases begin to increase by mid to late July. 
However, if by then the vaccine coverage is extensive, and vaccines prove to be between 75\% to 95\% effective,
this wave will be barely noticeable even till September.
\item 
From Figure 1 it may appear that vaccines are less impactful then one may naively imagine.
A key reason for that is that in a city where 80\% of the population has already been exposed to the virus, 
only a small fraction (say, around 20\%) of the vaccines are administered to the susceptible population.
 \end{itemize}

 Below we also plot $R_t$, the infection growth rate at time $t$, and highlight some interesting observations.

\section{Details}

\begin{figure}
      \centering
     \includegraphics[width=\linewidth]{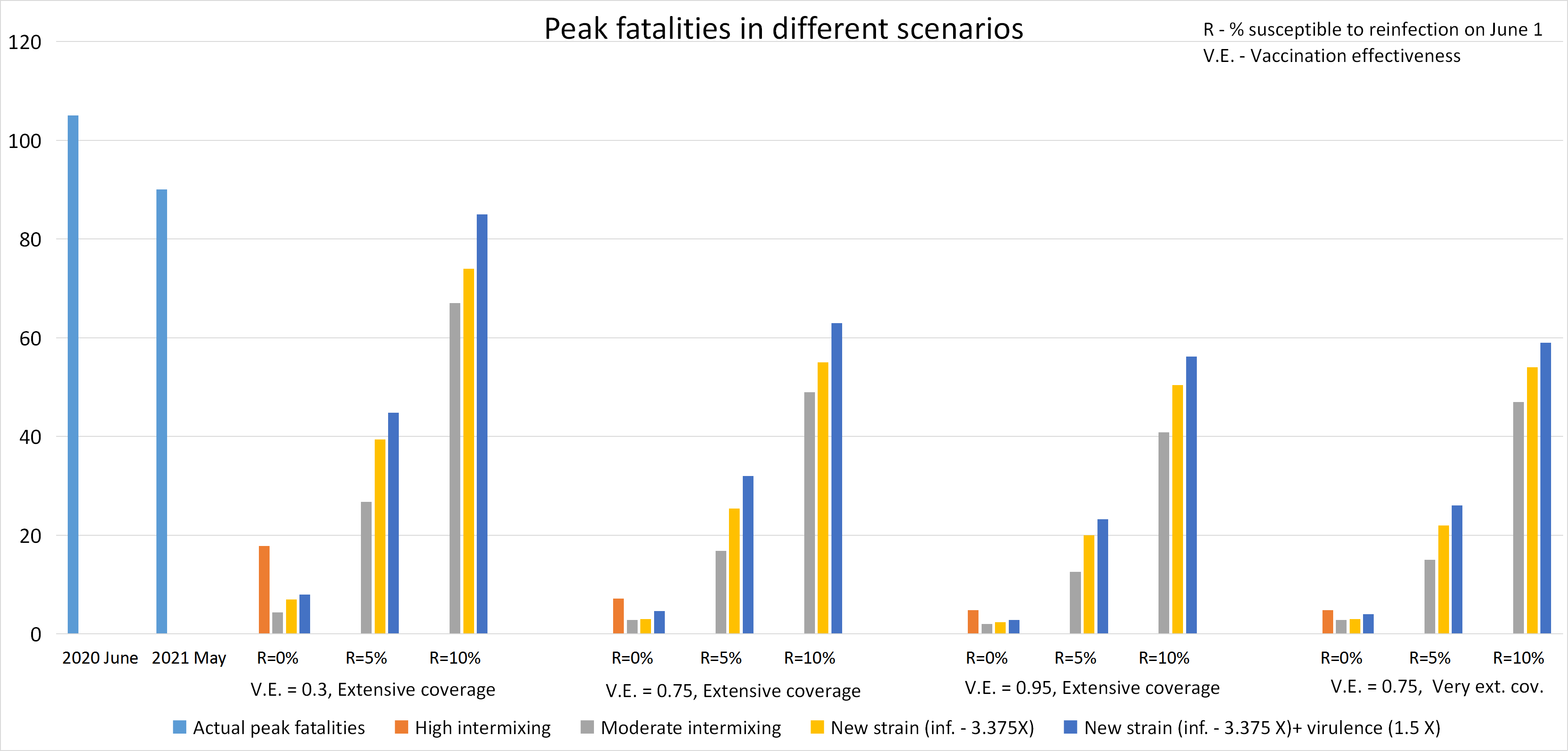}
      \caption{\small 
      Peak fatality numbers under scenarios corresponding to varying reinfection rates, vaccine effectiveness and coverage,
      presence of infectious and virulent variant (in addition to the delta variant).
    Under the high intermixing orange bar scenario, we assume that that the city is opened up  at 60\% level in June, 80\% in July and 100\% thereafter.
The compliance is 20\% in non-slums and 10\% in slums in June. This reduces to 10\% and zero July onwards. 
In all other scenarios we assume that the city is opened up  at 60\% level, and the compliance  
 is 20\% in non-slums and 10\% in slums June onwards.
 } \label{peak_fatalities}
  \end{figure}

\begin{figure}
      \centering
     \includegraphics[width=\linewidth]{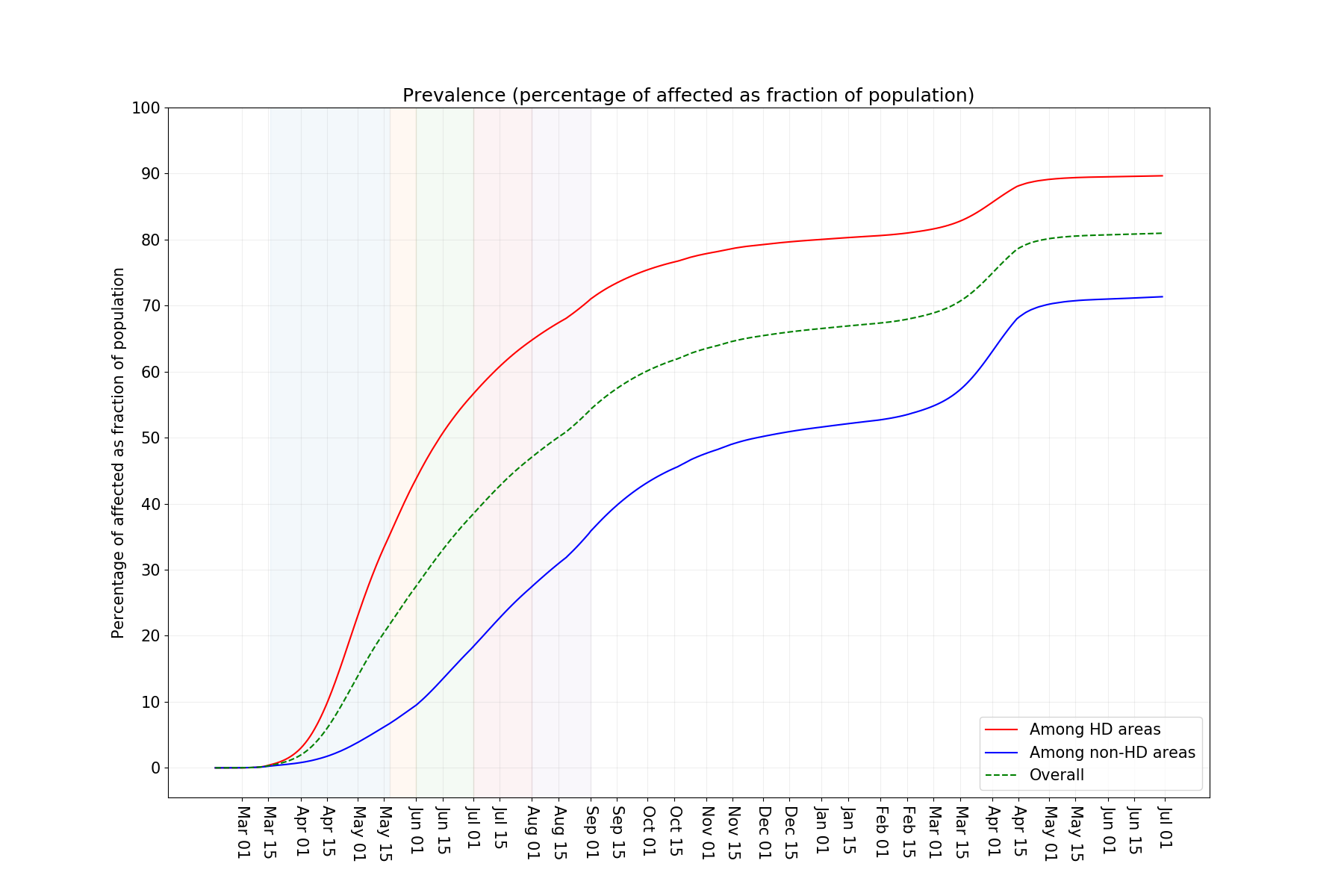}
      \caption{\small 
      Prevalence (percentage of exposed as a fraction of the population). No new strain apart from the delta strain. No reinfections. City opens at 60\% in June 2021, 80\% in July and 100\% from August. Compliance is 0.2 (non-slums), 0.1 (slums or high density HD areas) in June and 0.1 (non-slums), 0 (slums) from July onwards.
 } \label{prevalence_affected_1}
  \end{figure}

\begin{figure}
      \centering
     \includegraphics[width=\linewidth]{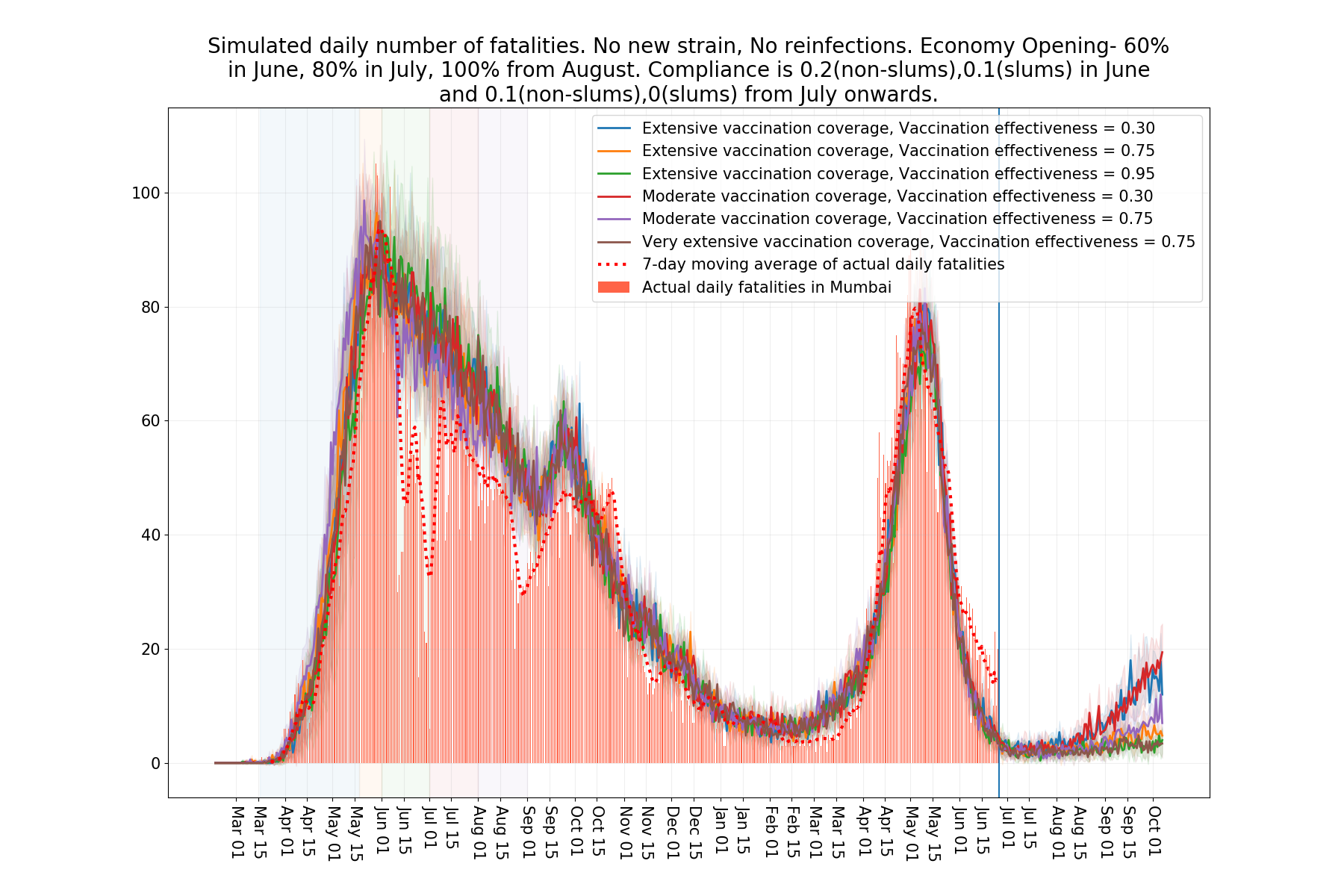}
      \caption{\small 
       Simulated daily number of fatalities. No new strain apart from the delta strain. No reinfections. 
     City opens at  60\% in June 2021, 80\% in July and 100\% from August. Compliance is 0.2 (non-slums), 0.1 (slums) in June and 0.1 (non-slums), 0 (slums) from July onwards.  
 } \label{daily_deaths_1}
  \end{figure}

Figure~\ref{peak_fatalities} illustrates the key summary of our analysis. We describe it in some detail. To set the stage some estimates our in order. As mentioned earlier, we estimate that around 65\% of Mumbai was infected on February 1, 2021 at the beginning of the  second wave (around 80\% in slums and a little over 50\% in non-slums. See Figure~\ref{prevalence_affected_1}. Our model assumes, as per census 2011, that 53\% of Mumbai resides in slums\cite{MCGM_census2011_report}). To best fit the fatality data from March to May 2021 where we assume that the delta variant accounted
for 2.5\% of the infected on February 1 and is 2.25 times more infectious than the earlier Wuhan variant in Mumbai. We further assume that the population infected by this variant, if it needs ICU treatment, it is for 18 days,  an increase of 10 
days over the earlier variant. This increase helps our model better fit the fatality data and is supported by our ongoing analysis of oxygen consumption data in Mumbai and Thane where we find that for the patients infected by the delta variant increasing the patient ICU stay by 8-10 days provides a better fit.
Vaccines are administered in our model to roughly match the Co-win statistics for Mumbai \cite{cowin_stat}.
This is discussed  in more detail in the appendix. In our earlier reports including in \cite{May_report_2021} and \cite{October_report_2020}, the symptomatic fraction of the exposed
population is kept at 40\%. We find that for the delta variant increasing it to 45\% (this leads to 12.5\% increased virulence)
provides a somewhat better fit to the fatality data.  Remaining
 parameters for the model are set as  in \cite{May_report_2021}.

Under this setting, we find that on June 1, 80\% of Mumbai is infected, 90\% in slums and 70\% in non-slums. A little more than 10\% of the susceptible have been vaccinated with a single dose (around 14 lakh administered a single dose 
and over 3 lakh administered the second dose  in March and April that we conservatively assume
 to become effective
a month later. We do not consider the vaccines administered to health care workers in our analysis).

\subsection{Scenario analysis of peak fatalities}
 
Given the large model uncertainty,  there is an element of curve fitting through parameters in our model especially 
after February 1, 2021. Even then it is reasonable to believe that while the absolute numbers that we observe may have large error bars,  our model output is useful in comparing scenarios. In this spirit, in Figure 1 we compare the reported fatalities peak in June 2020, and in May 2021 and compare them to 40 scenarios constructed as follows:
 \begin{itemize}
 \item
 Four scenarios are related to 
 vaccine administration. These closely match the vaccines administered in Mumbai till end May. In three of these scenarios we assume that there is extensive coverage but the vaccine effectiveness takes values
 0.3, 0.75 and 0.95. Under extensive coverage, monthly distribution
 of vaccines from March to September 2021 is 5 lakh, 12.2 lakh, 7.5 lakh, 20 lakh, 30 lakh, 30 lakh and  30 lakh doses.
 Again details are provided in the appendix.  We make a somewhat conservative assumption that  vaccines given in any month become effective a month later.
Also,  vaccine effectiveness of x means that every person receiving a dose
is moved to the fully recovered category with probability x.
This is implemented by assuming that such a person is selected
with probability x. However, it becomes recovered with probability 0.5 after the first dose. 
If it does not recover after first dose, and is not infected by
the time the second dose is administered, then it recovers with probability 1 after the second dose. Under the fourth scenario we assume
very extensive coverage (reaching 2 lac doses a day in July) and vaccine effectiveness of 0.75.

\item
For each of the four vaccination scenarios, we consider three reinfection scenarios. $R=0$, $R=5\%$ and $R=10\%$.
These increase the pool of susceptible population  by assuming that fraction R of the recovered become
susceptible.  So with 20\% susceptible population 
on June 1, $R=10\%$ implies that 8\% of the overall population now joins 
the susceptible pool increasing it to 28\%.  We further assume
that each recovered person who joins the susceptible pool and is again infected, faces the same disease progression dynamics as  a person who is  infected for the first time.
In practice, reinfections maybe milder than first time infections. So in essence we are only considering those
reinfections that are statistically as severe as the first time infections.
 Since it's not clear how reinfections will occur, as a worst case we assume that all of these occur on June 1.
This would show an early and higher peak than if we had assumed that reinfections are more spread out in the coming months.

Also, we have assumed zero reinfections up till end May 2021 in our model, and with the selected parameters,
the model matches the fatalities well. Thus, if in  reality  there were reinfections that resulted in fatalities
by end May, they are implicitly captured by the model. Thus, reinfections introduced by setting $R>0$,
are over and above the past pattern of possibly fatal  reinfections.

The $R>0$ scenarios also provide insights for the case where the actual number of susceptible on June 1 are higher than the 20\% suggested by our model.

\item
For each vaccine administration and reinfection scenario, we also consider
three cases:
\begin{enumerate}
\item
There is no new strain over and above the dominant delta strain.
\item
There is a new strain that is present in small amount on June 1 (2.5\% of the infected), whose infectivity is 50\% higher than the delta strain
and 3.375 times that of the strain active in Mumbai last year. Again, June 1 is chosen because this would lead to an earlier and higher peak than if the new variant
were to evolve later. Virulence of this strain is assumed to be 
same as that of the delta strain.
\item
 On top of the infectiousness as above in the new strain, we assume that it is 50\% more deadly.
 \end{enumerate}
 \end{itemize}
 
 Our broad conclusions related to the scenario analysis of peak fatalities have already been discussed in the initial
 summary.
 
 In Figure~\ref{daily_deaths_1} we display the fatality projections corresponding to 
 a plausible scenario
  where the reinfections are mild so that they affect the fatality figures negligibly and 
 where the new variants (beyond the existing delta variant) have little impact.
 This corresponds to the first orange colored (high intermixing) bar in Figure 1.  
Here we assume that that the city is opened up  at 60\% level in June, 80\% in July and 100\% thereafter.
The compliance is 20\% in non-slums and 10\% in slums in June. This reduces to 10\% and zero July onwards. 
These numbers reflect the increase in laxness that Mumbai may observe as the cases and fatalities continue their decline.
 In this scenario, as we mentioned earlier in the initial summary, 
 the city may observe a small wave where the fatalities begin to increase by around mid-August. Hence, since the fatalities lag reported cases, the cases may begin to increase by mid to late July. 
 
Fortunately, as is clear from Figure~\ref{daily_deaths_1},
 if in the months of June, July and August the vaccine coverage is extensive, and vaccines prove to be between 75\% to 95\% effective, this wave will be barely noticeable even by September.

\subsubsection{Modelling $R_t$}
\begin{figure}
      \centering
     \includegraphics[width=\linewidth]{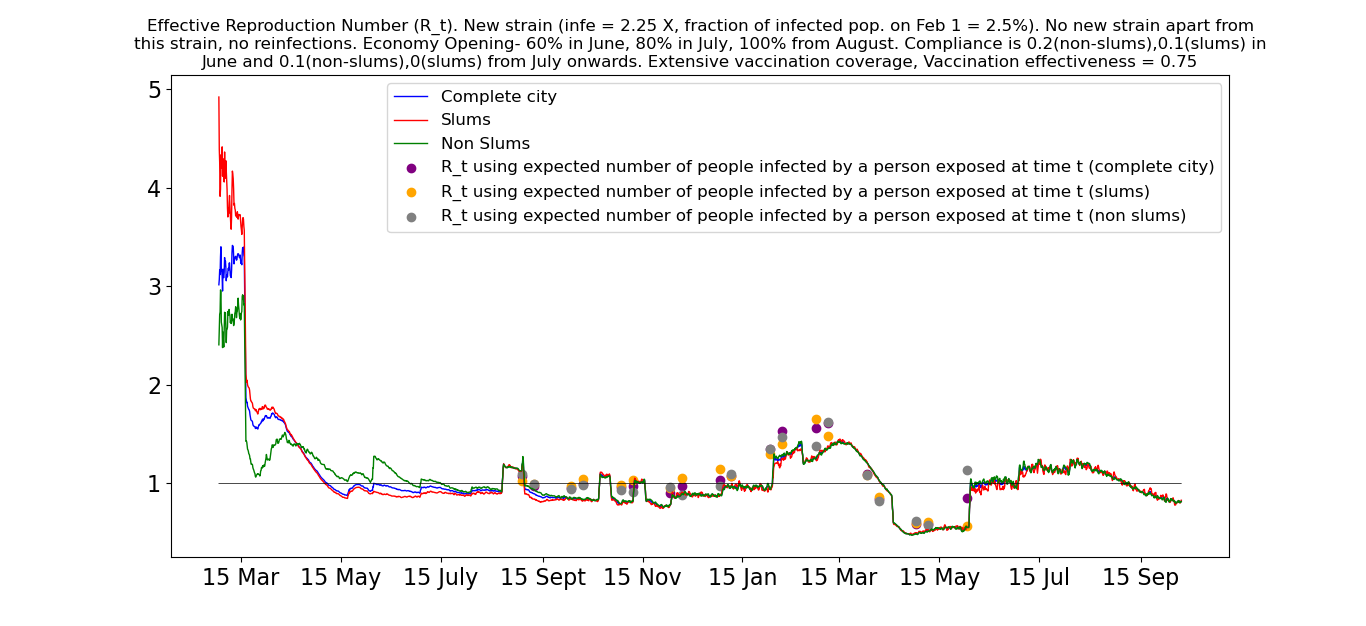}
      \caption{\small 
      Effective reproduction number ($R_t$) for complete city as well as for slums and non-slums. Delta strain (infectiousness = 2.25 X, fraction of infected pop. on Feb 1 = 2.5\%). In this scenario
      no new strain exists apart from the delta variant and there are no reinfections. The city opens at 60\% in June, 80\% in July and 100\% from August. Compliance is 0.2 (non-slums) and 0.1 (slums) in June and 0.1 (non-slums) and 0 (slums) from July onwards. We consider extensive vaccination coverage scenario with  vaccination effectiveness = 0.75. 
 } \label{R_t_all}
 
  \end{figure}

 In Figure~\ref{R_t_all}, we plot $R_t$ in the above scenario
as per our model. Recall that $R_t$ is designed to measure  the growth rate of the disease at time $t$.
$R_t>1$ signifies increasing infection in the system and $R_t<1$ signifies declining infection.
We define $R_t$ in two different ways. In th`e first method, it equals the ratio
of the average number exposed between time $t$ and $t+1$ divided by the 
average number of people who stop being infective (in our model this equals 
the sum of the people who recover from
the  infectious symptomatic state and those
who stop being infectious to others because they are transferred to hospitals) in this time period.

Interestingly, as can be seen in Figure~\ref{R_t_all}, this number becomes roughly equal
for slums and non-slums in Mumbai in August 2020, and more or less stays equal thereafter.
The rationale for this may be that the interactions between the two populations in communities as well as
in local trains, tends to move the two groups to equalize the differences. 

Another way to define $R_t$ is to track the average of the  number of infections caused by an individual
exposed at time $t$. These are measured at a few times and plotted as dots in Figure~\ref{R_t_all}.
Interestingly, this definition gives very similar numbers to the earlier one, and again these numbers are quite close for both slums and non-slums. Observe  that this is true even though prevalence between slums and non-slums is substantially different.

\section{Appendix}
\label{section:append}

In Section~\ref{sec:vaccine}  below we  discuss the underlying assumptions 
based upon the data from \cite{cowin_stat}, and outline the  vaccination schedule implemented 
in our model
in all the scenarios considered. 
Thereafter in Section~\ref{sec:fatalities}, we show some of the graphs of infection spread and fatalities corresponding to 
the 40 scenarios that we had outlined earlier in the report.
In Section~\ref{sec:R_t}, we display the evolution of $R_t$ under some of the scenarios 
discussed earlier.

 \subsection{Vaccination schedule for the scenarios considered in 3rd wave analysis}
 \label{sec:vaccine}
 
For the vaccination coverage up to end of May 2021 we rely on the data from \cite{cowin_stat}. We make following
additional  assumptions:
\begin{itemize}
\item
In Mumbai the fraction of Covishield vaccine is  high (almost 93\%) \cite{cowin_stat}. So, for simplicity,  we assume that a single vaccine is being administered to Mumbai population.
\item 
We assume that initial vaccine doses were mainly administered to healthcare workers.  As we do not model healthcare workers in our simulator separately, we do not take into account the first dose vaccines administered before March 1, 2021 and second doses before Apr 1, 2021.
\item
The elderly (above 60 years of age) were given the first vaccination dose in Mumbai from March 1 on wards. These were made available to 45 years and older from April 1 and for people  between 18-45 years, vaccines were made available from May 1. Any numbers for these age categories before the respective starting dates are not taken into account in our simulations. 
\item
As mentioned earlier,  we assume that  the above vaccine doses become effective  1 month later in our model.
(These are administered a month later in our model and are assumed to be immediately effective). 
\item 
Second dose is administered within 4-8 weeks (1-2 months) for the people vaccinated in April in our model (those
who were actually administered the vaccine in March)  while those vaccinated after April 31 (administered the vaccine 
 after March 31) will receive the second dose after 12 weeks (appx. 3 months) of the first dose as per the notification of Government of India \cite{covid_second_dose}. 
\end{itemize}

\begin{figure}
      \centering
     \includegraphics[width=\linewidth]{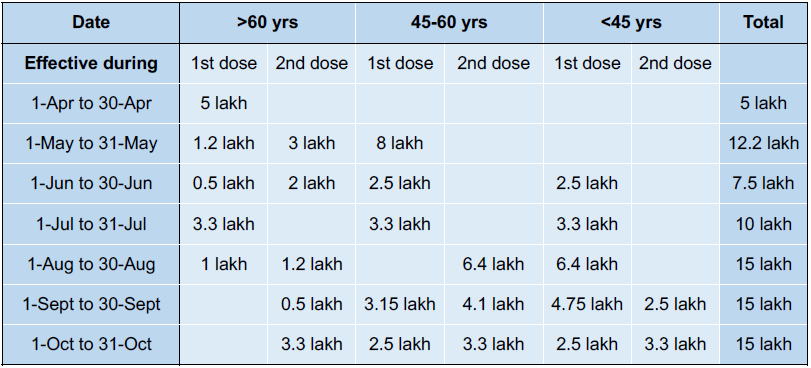}
      \caption{\small 
     Moderate vaccination coverage
 } \label{moderate}
  \end{figure}

\begin{figure}
      \centering
     \includegraphics[width=\linewidth]{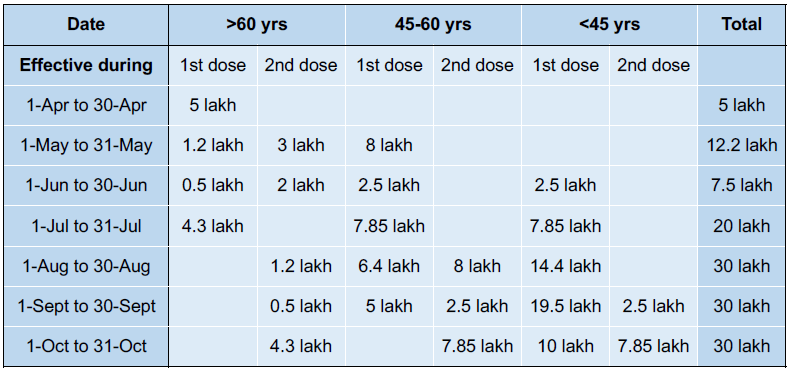}
      \caption{\small 
      Extensive vaccination coverage
 } \label{extensive}
  \end{figure}

\begin{figure}
      \centering
     \includegraphics[width=\linewidth]{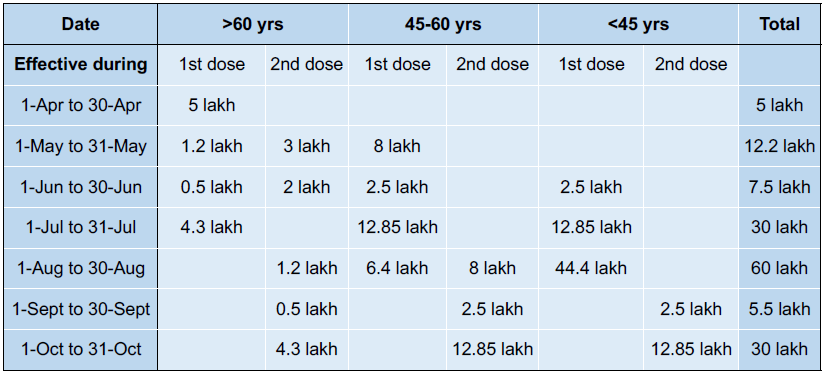}
      \caption{\small 
     Very extensive vaccination coverage
 } \label{very_extensive}
  \end{figure}

We consider following 3 different vaccination coverage scenarios. Detailed numbers for the scenarios are given in respective figures. (Quarter of these numbers  are vaccinated during the specified month at four equally spaced instances) :
\begin{itemize}
    \item 
    Moderate vaccination coverage. See Fig \ref{moderate}.
    \item 
    Extensive vaccination coverage See Fig \ref{extensive}.
    \item
    Very extensive vaccination coverage. See Fig \ref{very_extensive}.
\end{itemize}

Following methodology was used to arrive at the vaccines administered  after the month of June:
\begin{itemize}
    \item 
    The proportion of vaccines in three age groups (18-44 yrs, 45-60 yrs, $>$60 yrs) for the month of June 2021 in our model (actually administered  in May 2021) was appx. 33\% \cite{cowin_stat}. We try to maintain these ratio in subsequent months, unless everyone in an age group has been vaccinated. 
    \item
    People for whom second dose is due get priority over the people of the same age group requiring the first dose (provided vaccine is available for their age group).
    \item 
    Approximate total population in different age groups is 13 lakh (above 60 yrs), 27 lakh (45-60 yrs), 60 lakh (18-44 yrs).
    \item
 We assume that approximately 80\%  people above 60 years will be vaccinated upto October, 2021.
 \end{itemize}

\subsection{Graphs associated with our scenario analysis}
 \label{sec:fatalities}

\begin{figure}
      \centering
     \includegraphics[width=\linewidth]{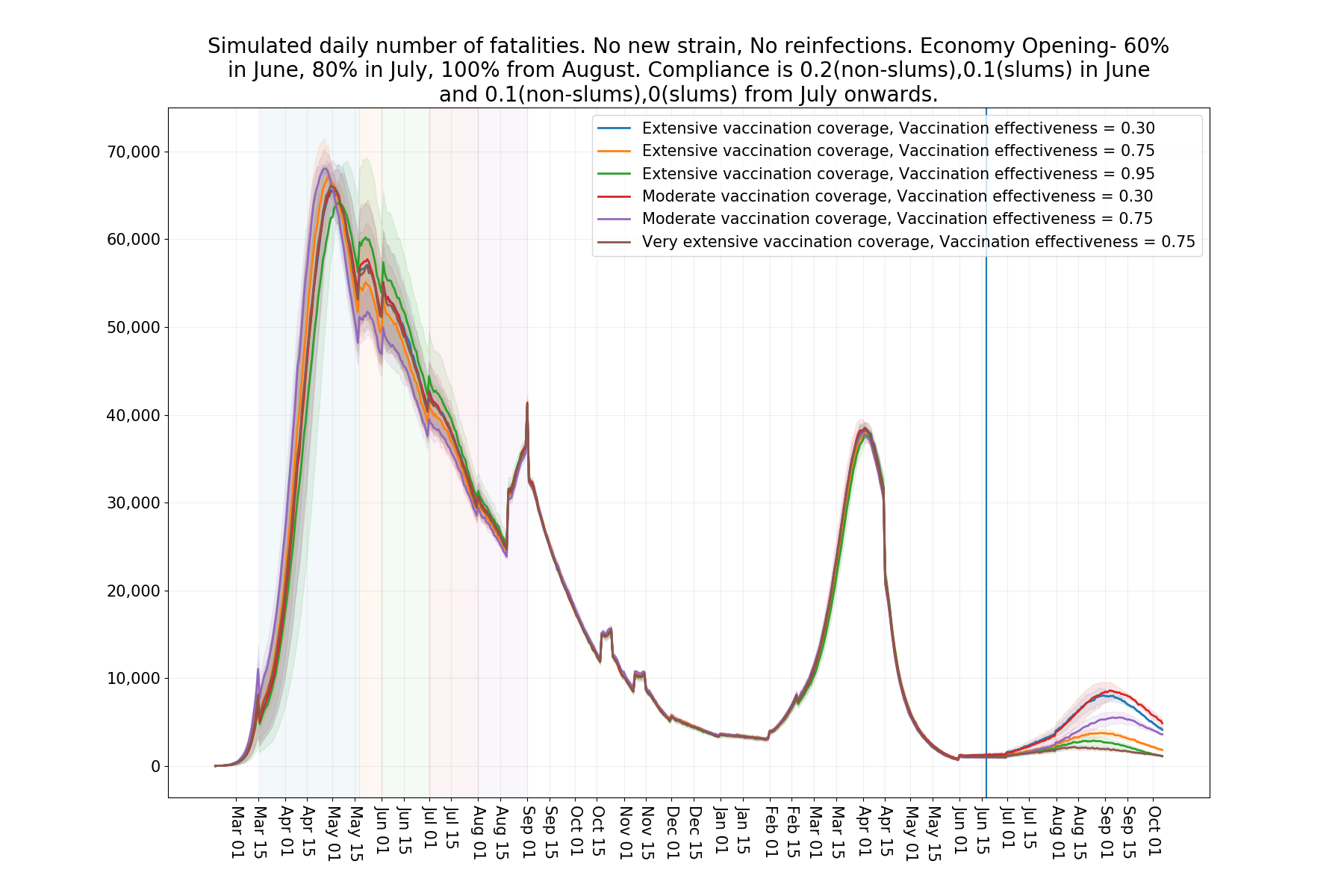}
      \caption{\small 
       Simulated daily number of new infections. No new strain apart from the delta strain. No reinfections. City  has  60\% mobility in June, 80\% in July and 100\% from August. Compliance is 0.2 (non-slums), 0.1 (slums) in June and 0.1 (non-slums), 0 (slums) from July onwards.
 } \label{daily_new_infections_1}
  \end{figure}

Recall that  in Figure~\ref{daily_deaths_1} we displayed the fatality projections under varying vaccination effectiveness and coverage corresponding to 
 a plausible scenario
  where the reinfections are mild so that they affect the fatality figures negligibly and 
 where the new variants (beyond the existing delta variant) have little impact.
 Here we had assumed that that the city is opened up  at 60\% level in June, 80\% in July and 100\% thereafter.
The compliance is 20\% in non-slums and 10\% in slums in June. This reduces to 10\% and zero July onwards.  Figure \ref{daily_new_infections_1} shows  the daily new infections
corresponding to the scenarios considered in Figure \ref{daily_deaths_1}. 
This figure brings out the increase in infections that may be observed due to low vaccine effectiveness or coverage.

\begin{figure}
      \centering
     \includegraphics[width=\linewidth]{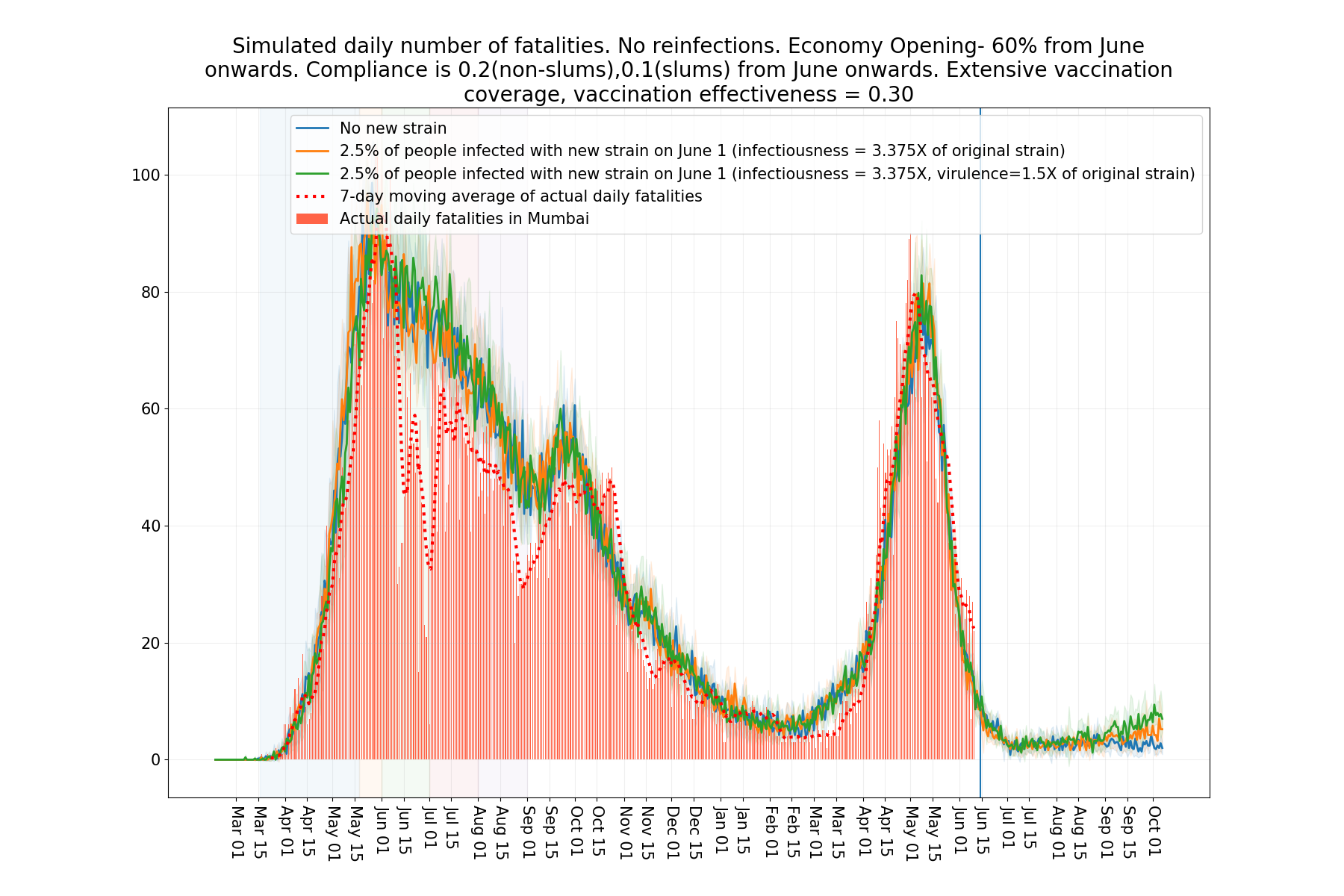}
      \caption{\small 
      Simulated daily number of fatalities. No reinfections. City  has  60\% mobility from June. Compliance is 0.2 (non-slums), 0.1 (slums) from June. Extensive vaccination (effectiveness=0.30) coverage.
 } \label{daily_deaths_2}
 
      \centering
     \includegraphics[width=\linewidth]{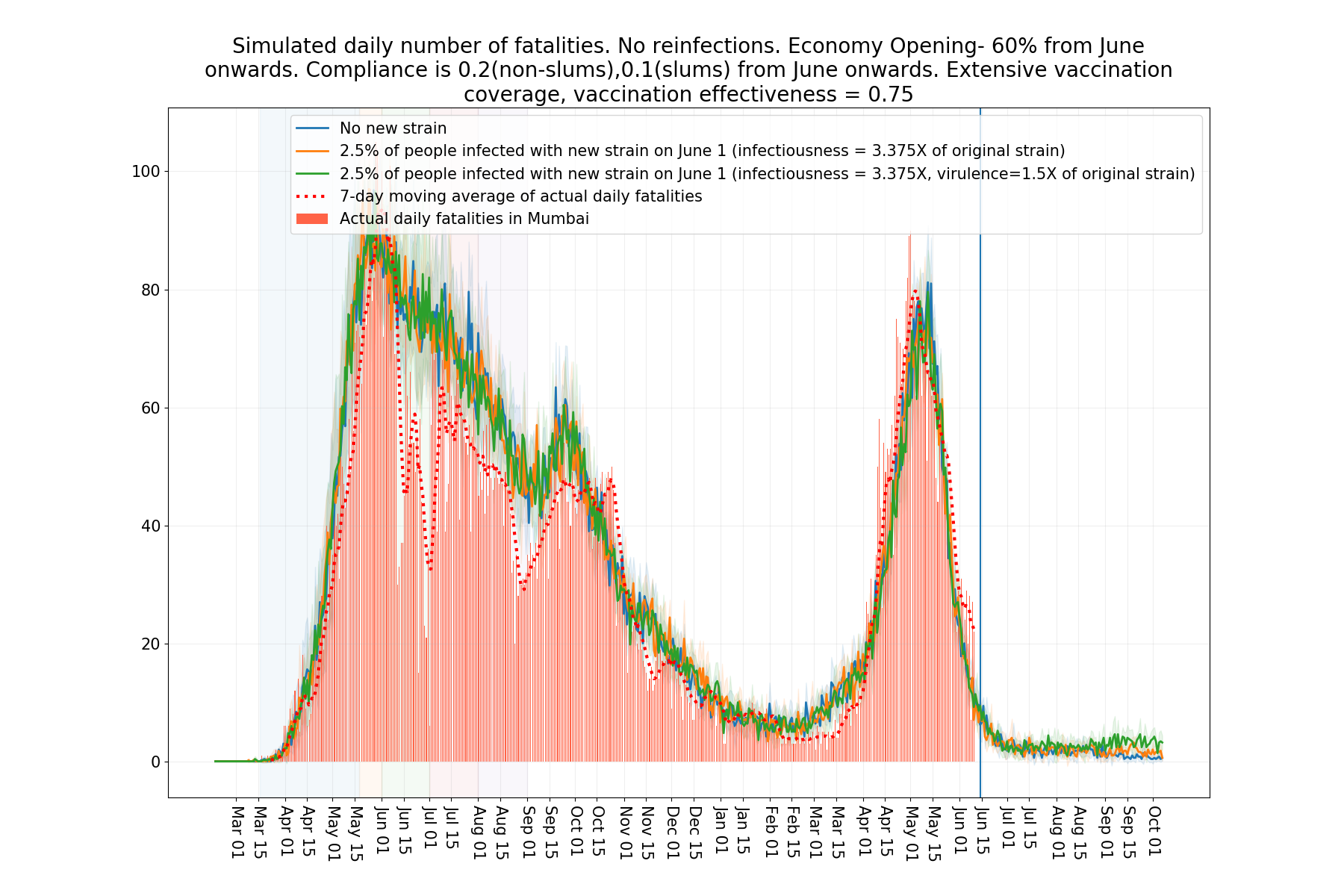}
      \caption{\small 
      Simulated daily number of fatalities. No reinfections. 
      City  has  60\% mobility from June. Compliance is 0.2 (non-slums), 0.1 (slums) from June. Extensive vaccination (effectiveness=0.75) coverage.
 } \label{daily_deaths_3}
  \end{figure}

\begin{figure}
      \centering
     \includegraphics[width=\linewidth]{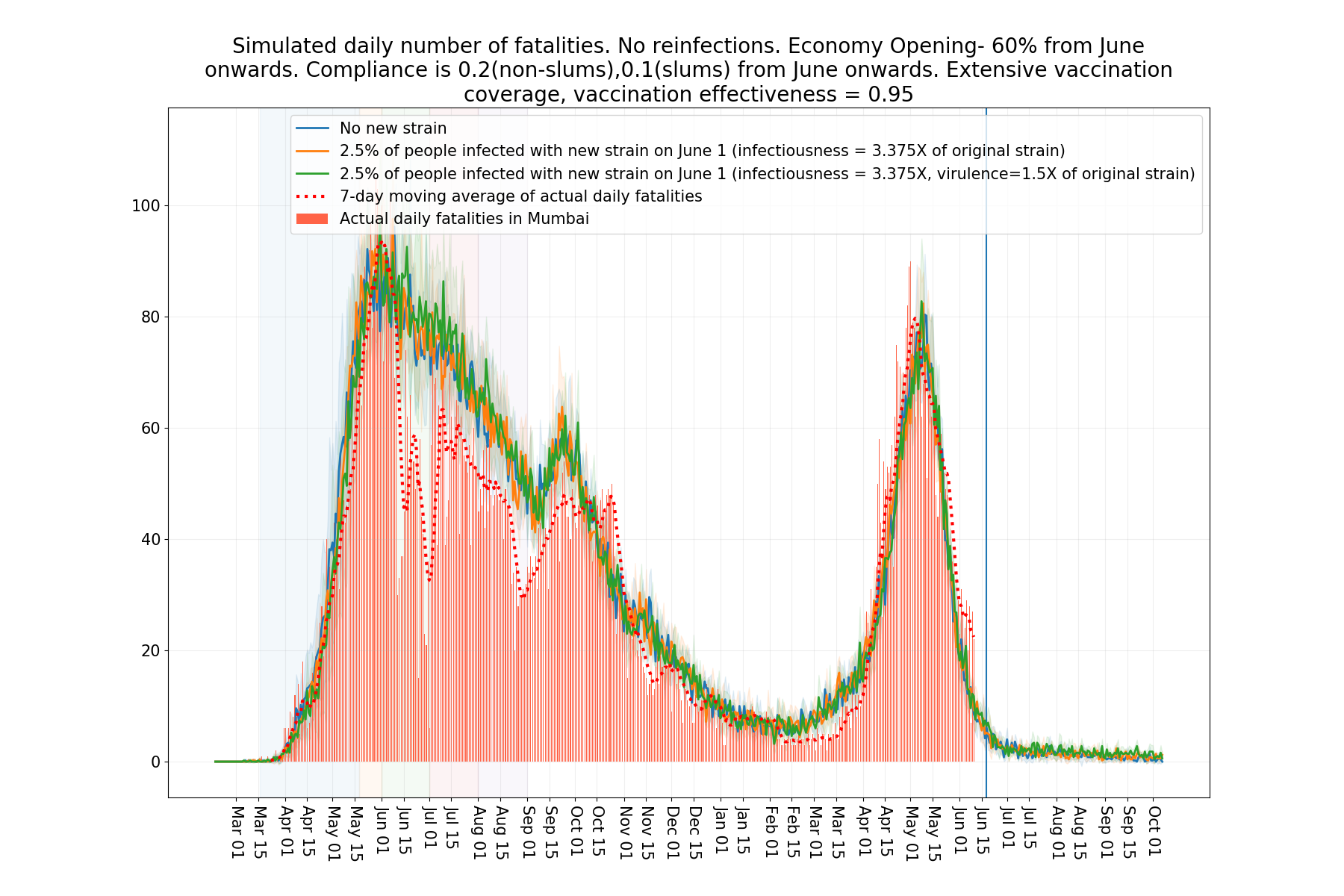}
      \caption{\small 
     Simulated daily number of fatalities. No reinfections. City  has  60\% mobility from June. Compliance is 0.2 (non-slums), 0.1 (slums) from June. Extensive vaccination (effectiveness=0.95) coverage.
 } \label{daily_deaths_4}
 
      \centering
     \includegraphics[width=\linewidth]{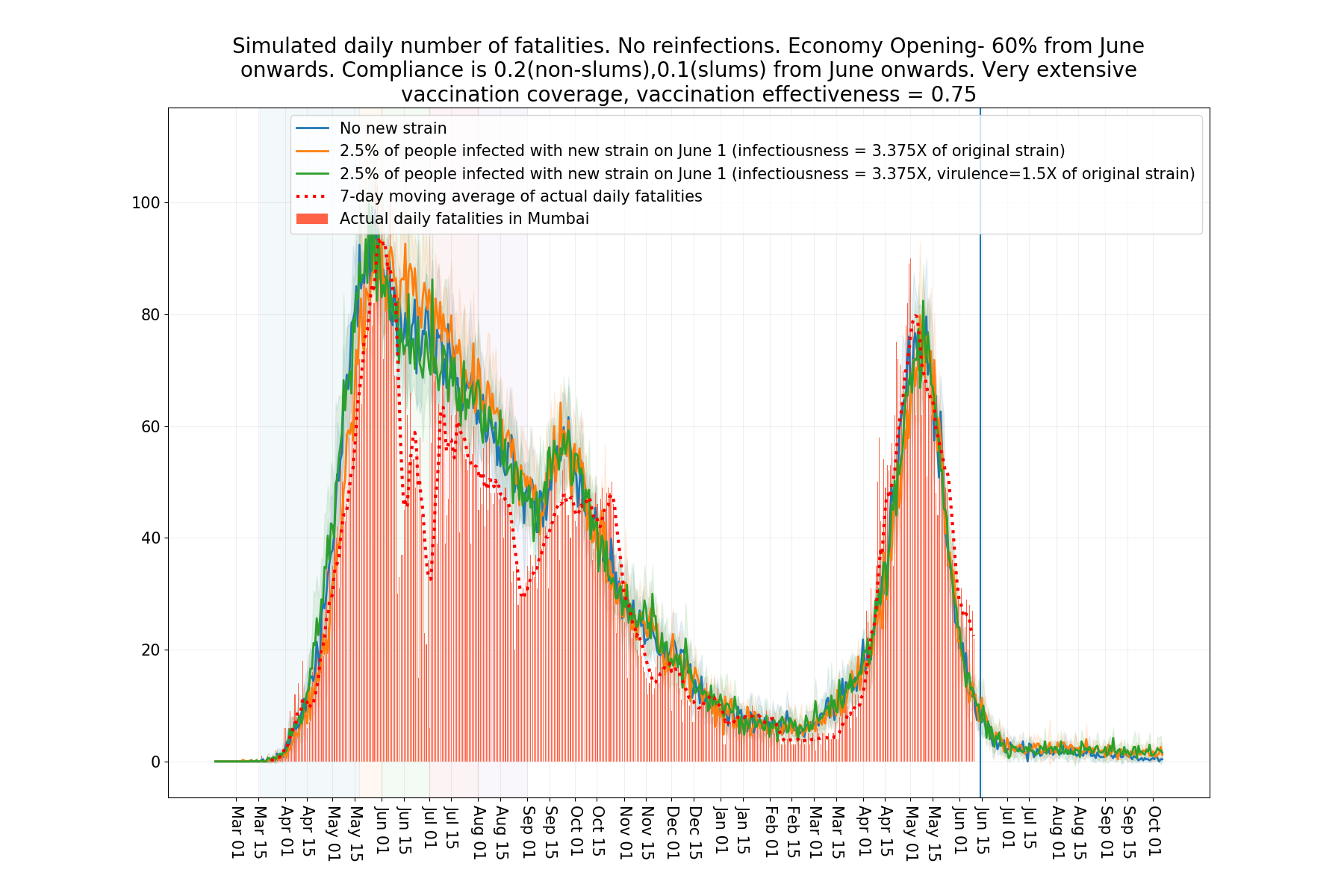}
      \caption{\small 
     Simulated daily number of fatalities. No reinfections. City  has  60\% mobility from June. Compliance is 0.2 (non-slums), 0.1 (slums) from June. Very extensive vaccination (effectiveness=0.75) coverage.
 } \label{daily_deaths_5}
  \end{figure}

\begin{figure}
      \centering
     \includegraphics[width=\linewidth]{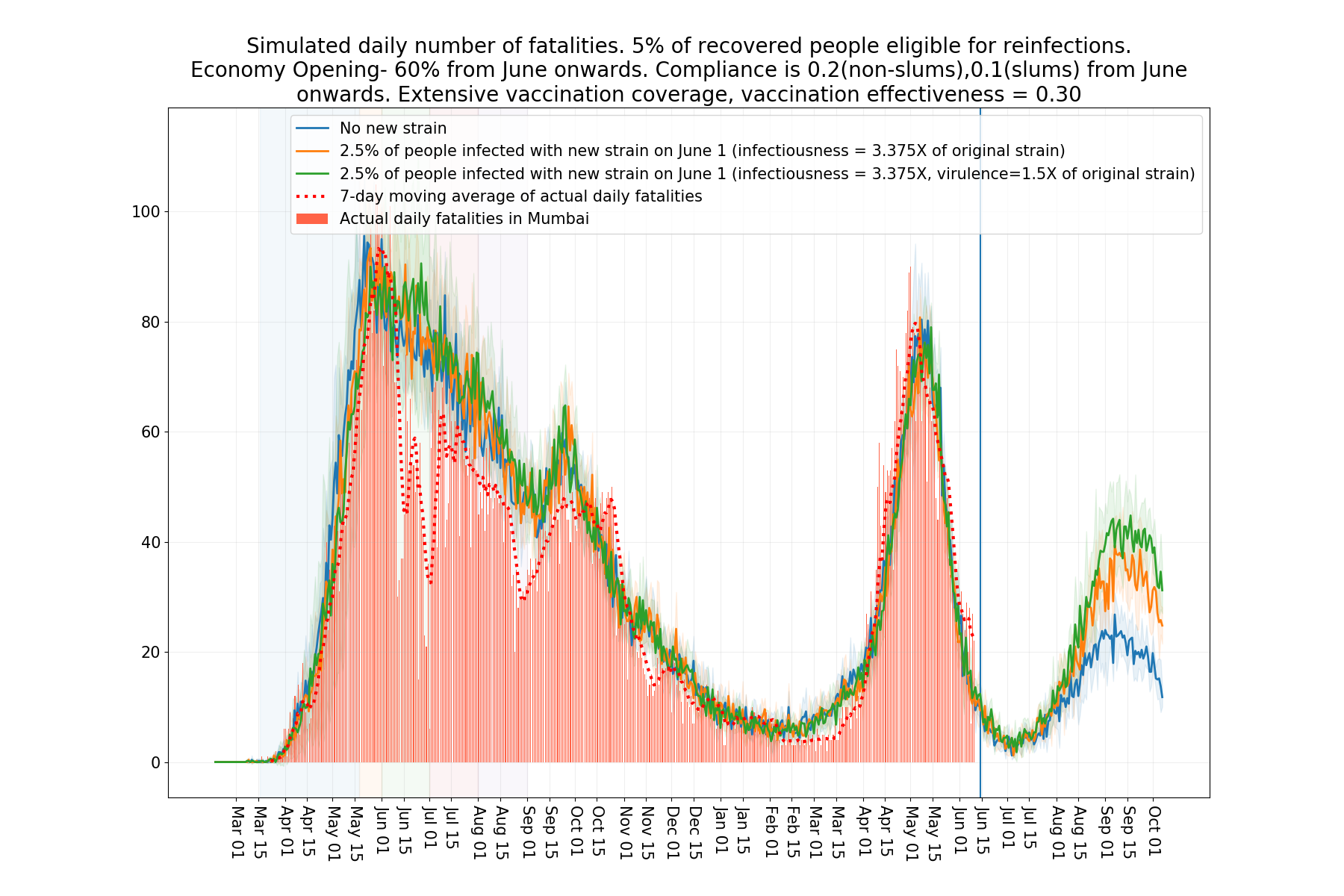}
      \caption{\small 
      Simulated daily number of fatalities. 5\% of recovered people eligible for reinfection on June 1. 
      City  has  60\% mobility from June. Compliance is 0.2 (non-slums), 0.1 (slums) from June. Extensive vaccination (effectiveness=0.30) coverage. 
 } \label{daily_deaths_6}
  
      \centering
     \includegraphics[width=\linewidth]{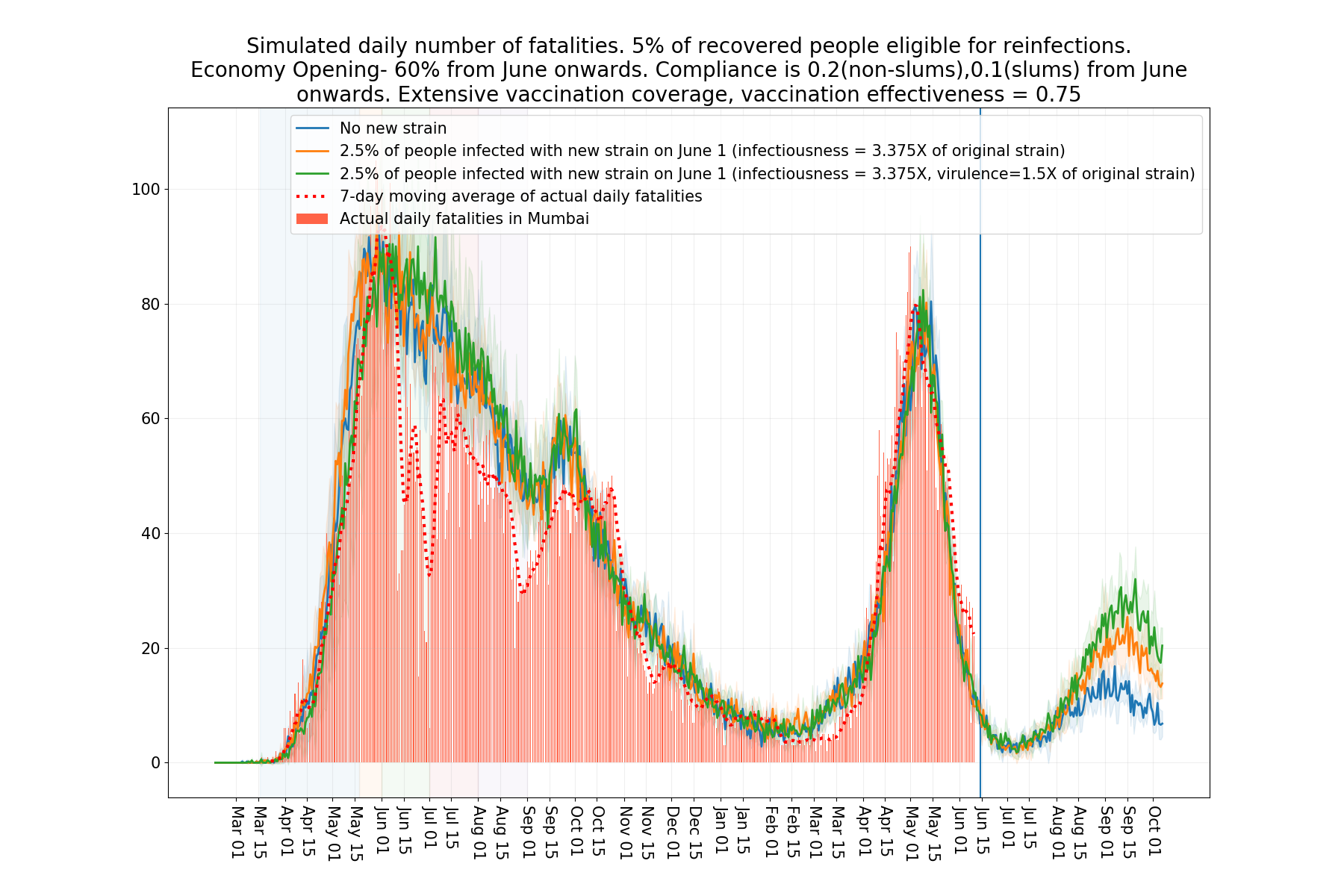}
      \caption{\small 
      Simulated daily number of fatalities. 5\% of recovered people eligible for reinfection on June 1. 
      City  has  60\% mobility from June. Compliance is 0.2 (non-slums), 0.1 (slums) from June. Extensive vaccination (effectiveness=0.75) coverage.
 } \label{daily_deaths_7}
  \end{figure}

\begin{figure}
      \centering
     \includegraphics[width=\linewidth]{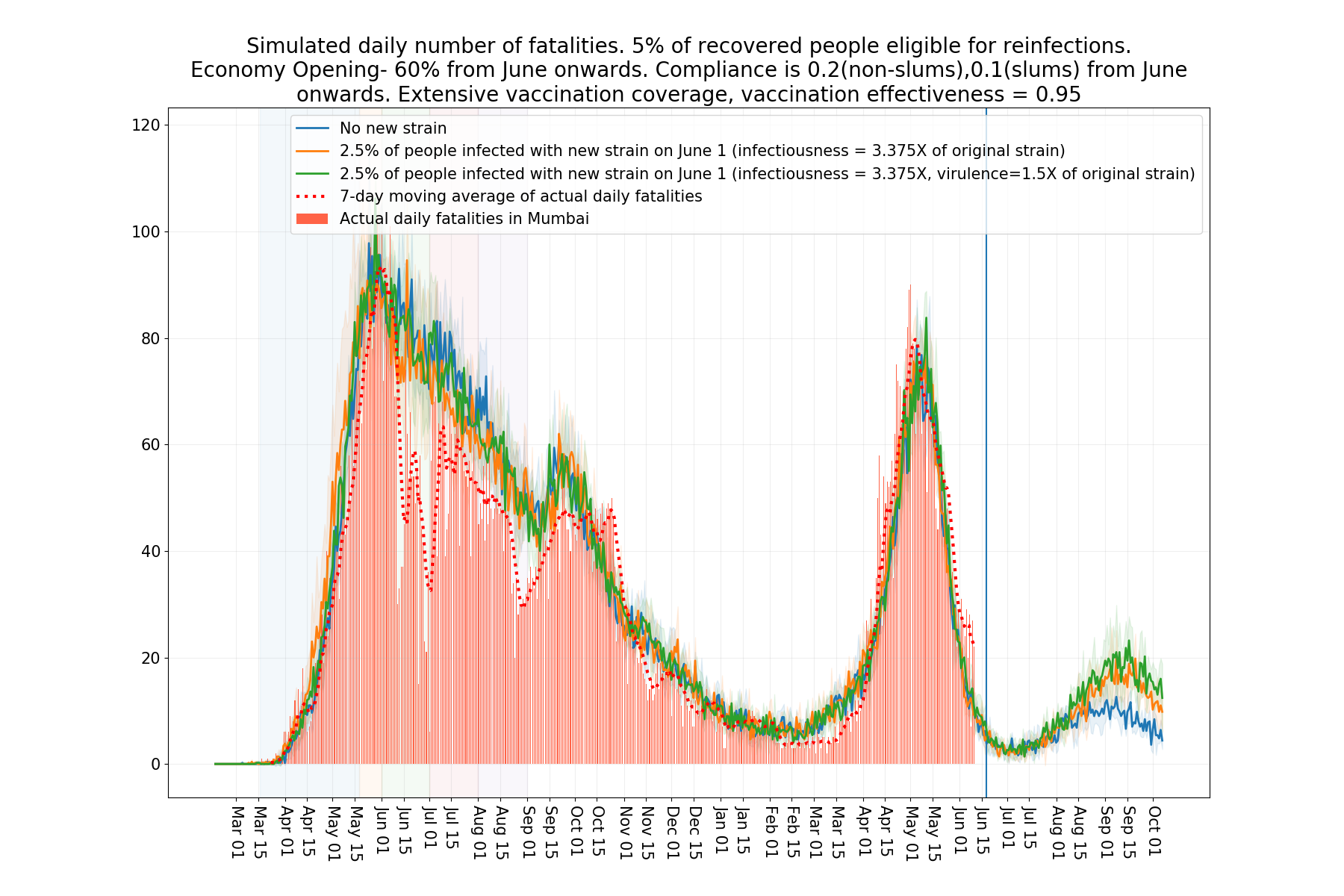}
      \caption{\small 
      Simulated daily number of fatalities. 5\% of recovered people eligible for reinfection on June 1. 
      City  has  60\% mobility from  June.  Compliance is 0.2 (non-slums), 0.1 (slums) from June. Extensive vaccination (effectiveness=0.95) coverage.
 } \label{daily_deaths_8}
  
      \centering
     \includegraphics[width=\linewidth]{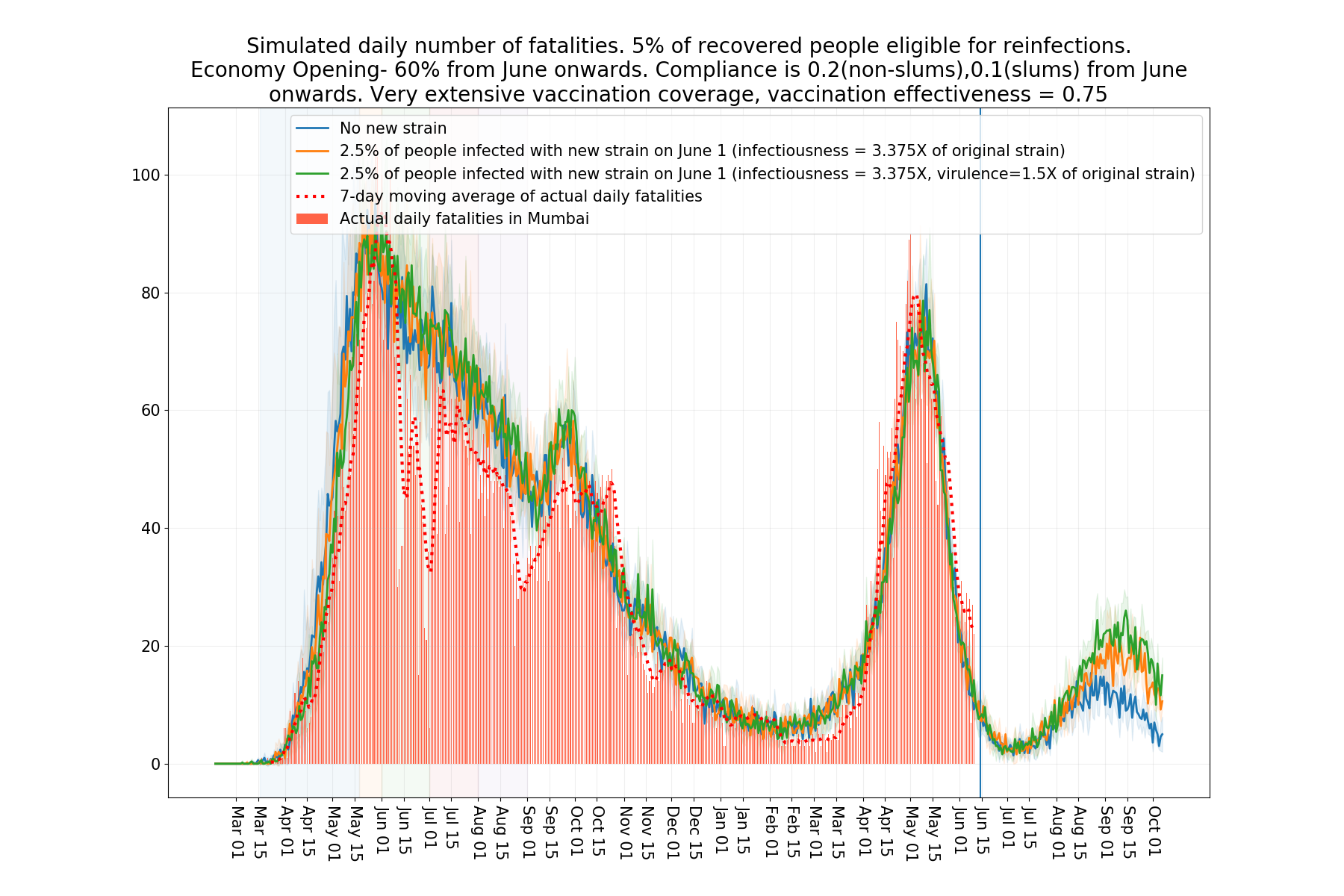}
      \caption{\small 
     Simulated daily number of fatalities. 5\% of recovered people eligible for reinfection on June 1. City  has  60\% mobility from  June.  Compliance is 0.2 (non-slums), 0.1 (slums) from June. Very extensive vaccination (effectiveness=0.75) coverage.
 } \label{daily_deaths_9}
  \end{figure}

\begin{figure}
      \centering
     \includegraphics[width=\linewidth]{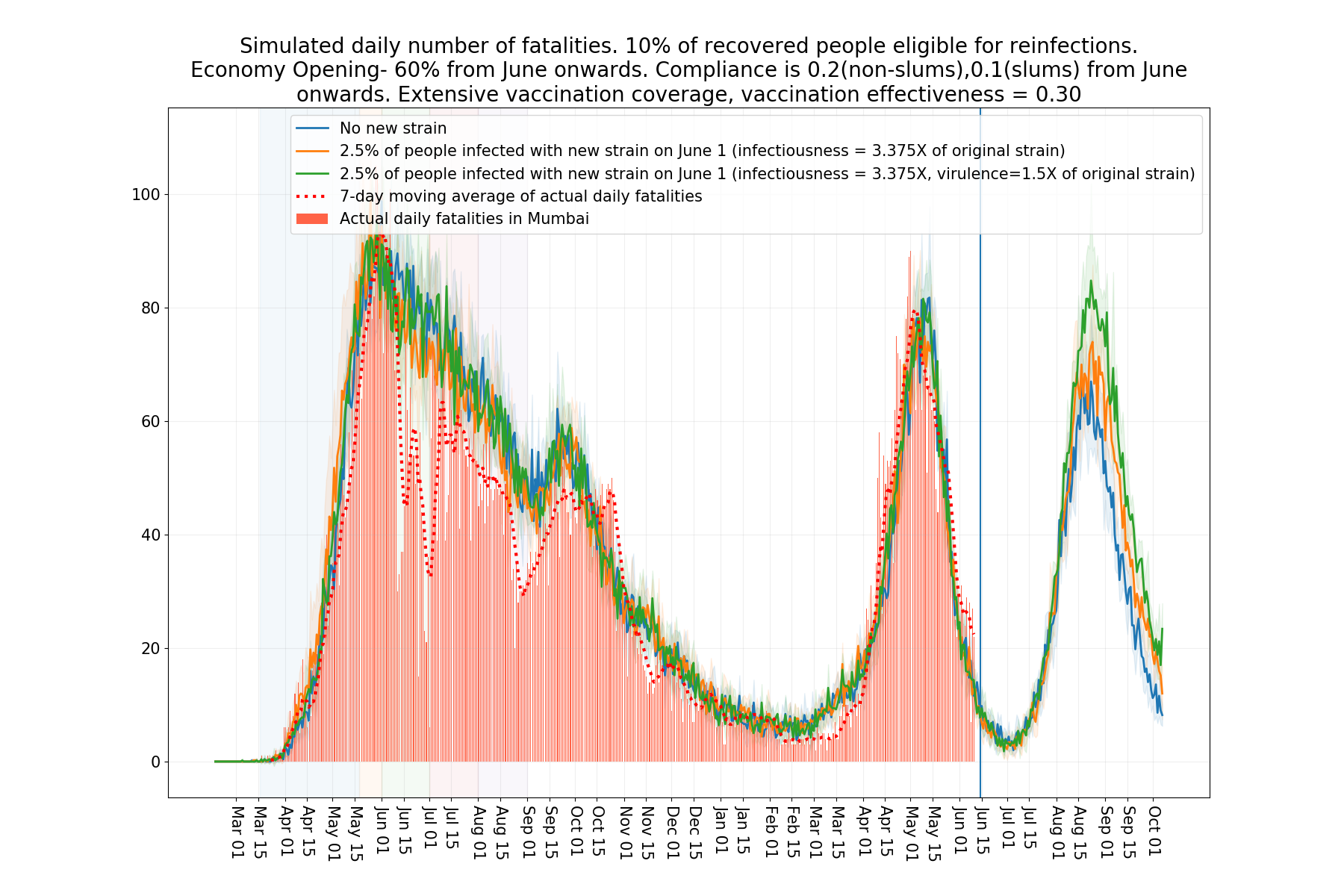}
      \caption{\small 
     Simulated daily number of fatalities. 10\% of recovered people eligible for reinfection on June 1. City  has  60\% mobility from  June.  Compliance is 0.2 (non-slums), 0.1 (slums) from June. Extensive vaccination (effectiveness=0.30) coverage.
 } \label{daily_deaths_10}
 
      \centering
     \includegraphics[width=\linewidth]{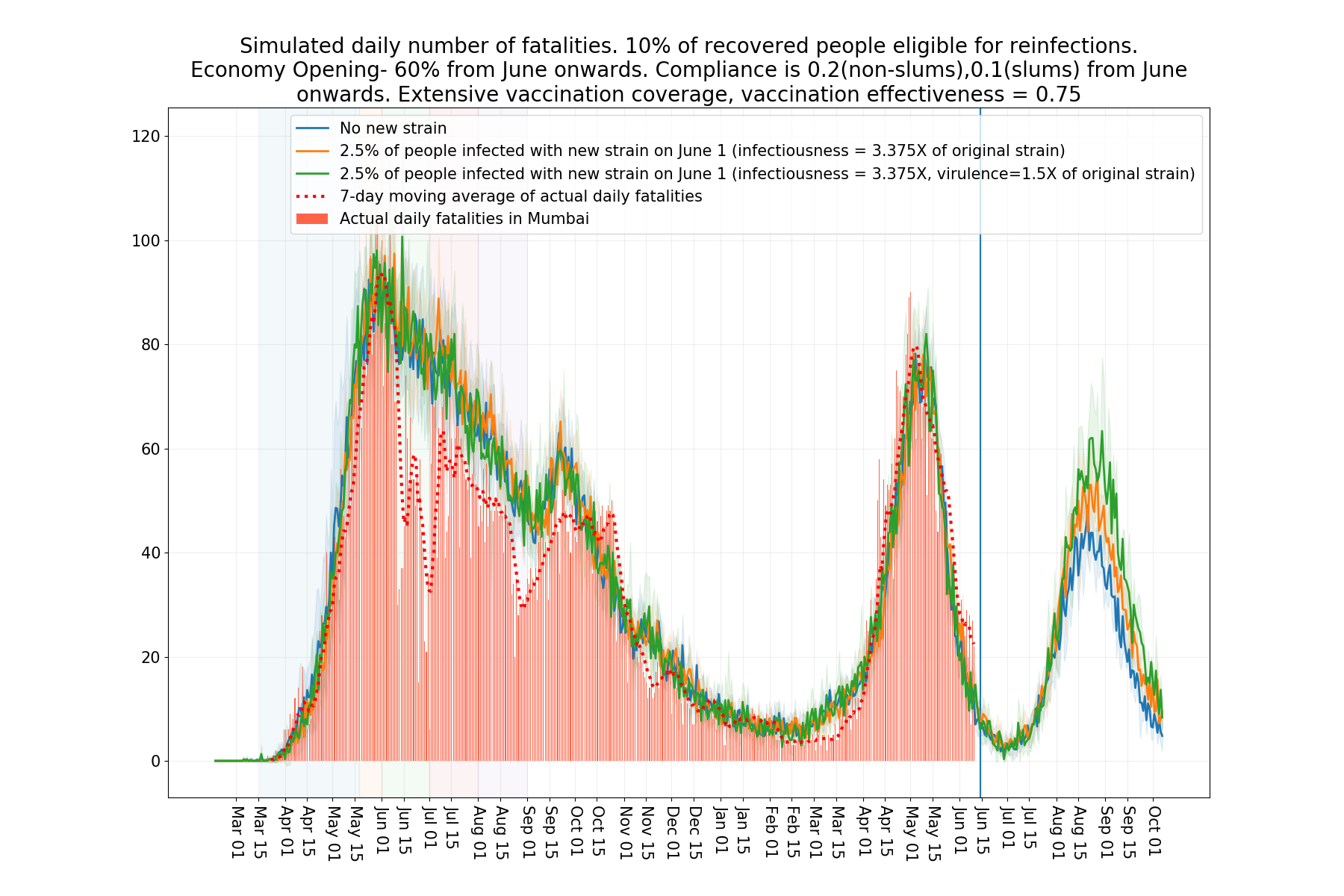}
      \caption{\small 
      Simulated daily number of fatalities. 10\% of recovered people eligible for reinfection on June 1. City  has  60\% mobility from  June.  Compliance is 0.2 (non-slums), 0.1 (slums) from June. Extensive vaccination (effectiveness=0.75) coverage.
 } \label{daily_deaths_11}
  \end{figure}

\begin{figure}
      \centering
     \includegraphics[width=\linewidth]{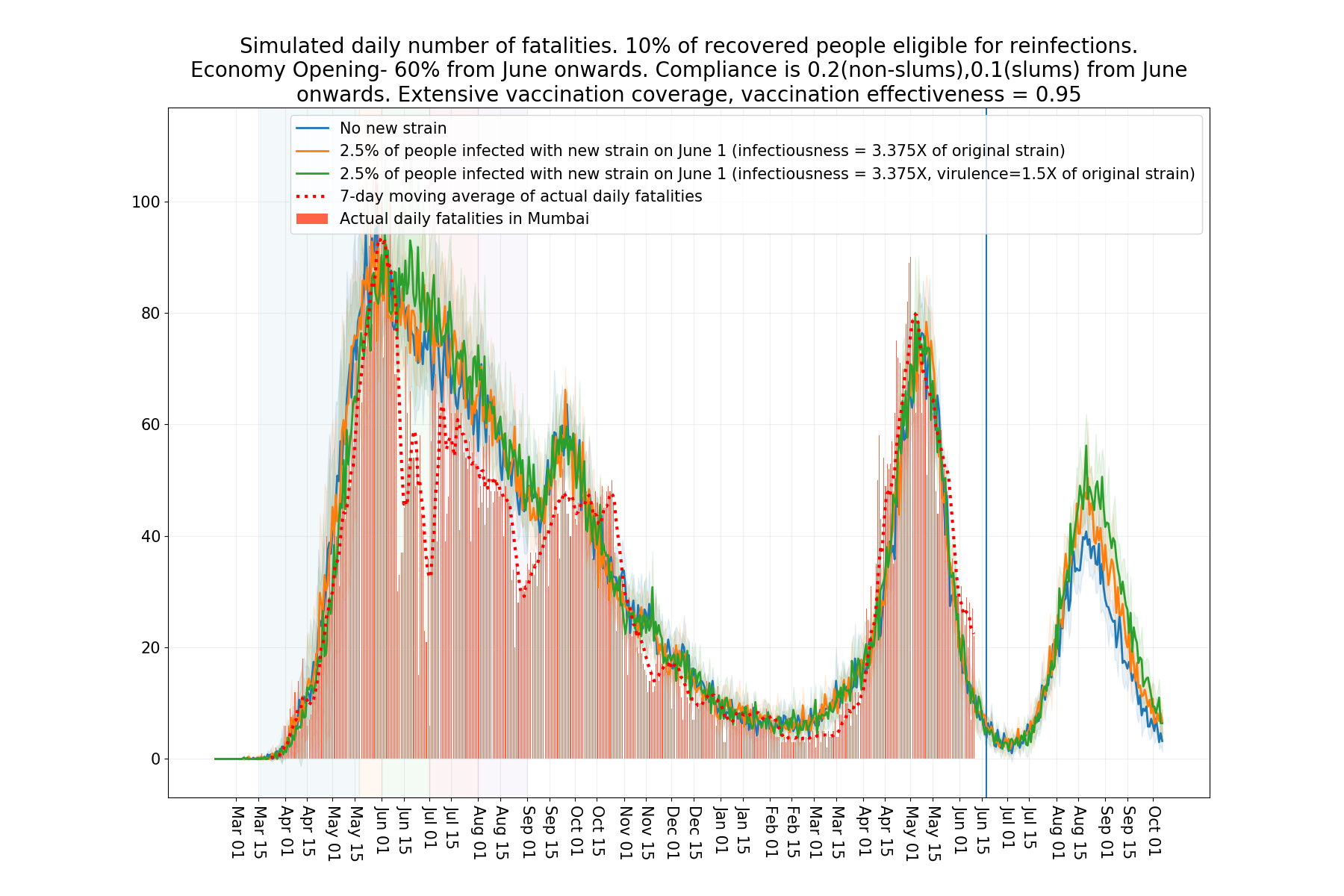}
      \caption{\small 
      Simulated daily number of fatalities. 10\% of recovered people eligible for reinfection on June 1. City  has  60\% mobility from  June.  Compliance is 0.2 (non-slums), 0.1 (slums) from June. Extensive vaccination (effectiveness=0.95) coverage. 
 } \label{daily_deaths_12}
 
      \centering
     \includegraphics[width=\linewidth]{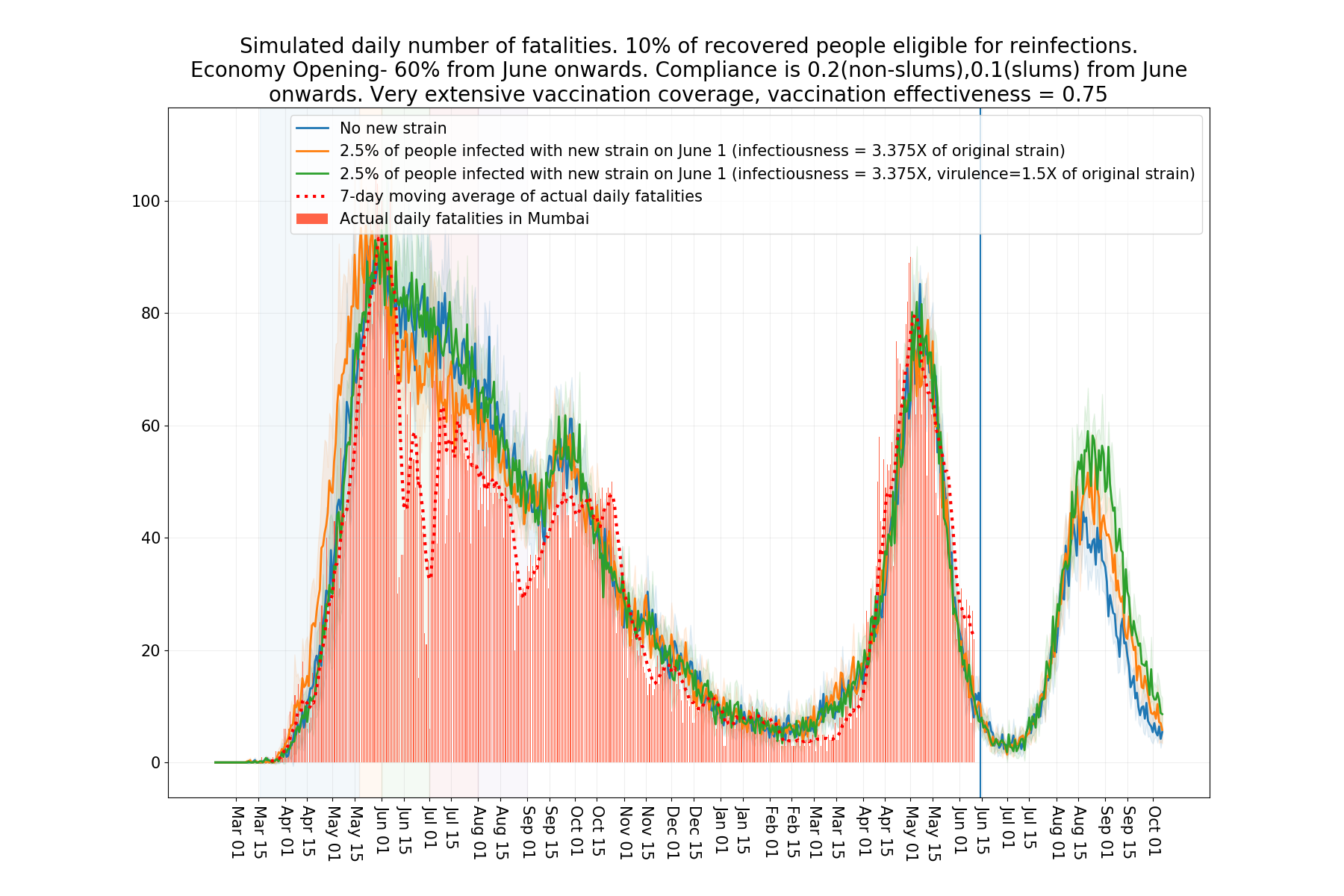}
      \caption{\small 
      Simulated daily number of fatalities. 10\% of recovered people eligible for reinfection on June 1. City  has  60\% mobility from  June.  Compliance is 0.2 (non-slums), 0.1 (slums) from June. Very extensive vaccination (effectiveness=0.75) coverage. 
 } \label{daily_deaths_13}
  \end{figure}

For completeness, below we show 
 the detailed daily fatalities graphs (see Figure \ref{daily_deaths_2} to \ref{daily_deaths_13}) for
 the different scenarios considered in Figure~\ref{peak_fatalities}.

\subsection{$R_t$ in some scenarios}
\label{sec:R_t}

\begin{figure}
      \centering
 
      \centering
     \includegraphics[width=\linewidth]{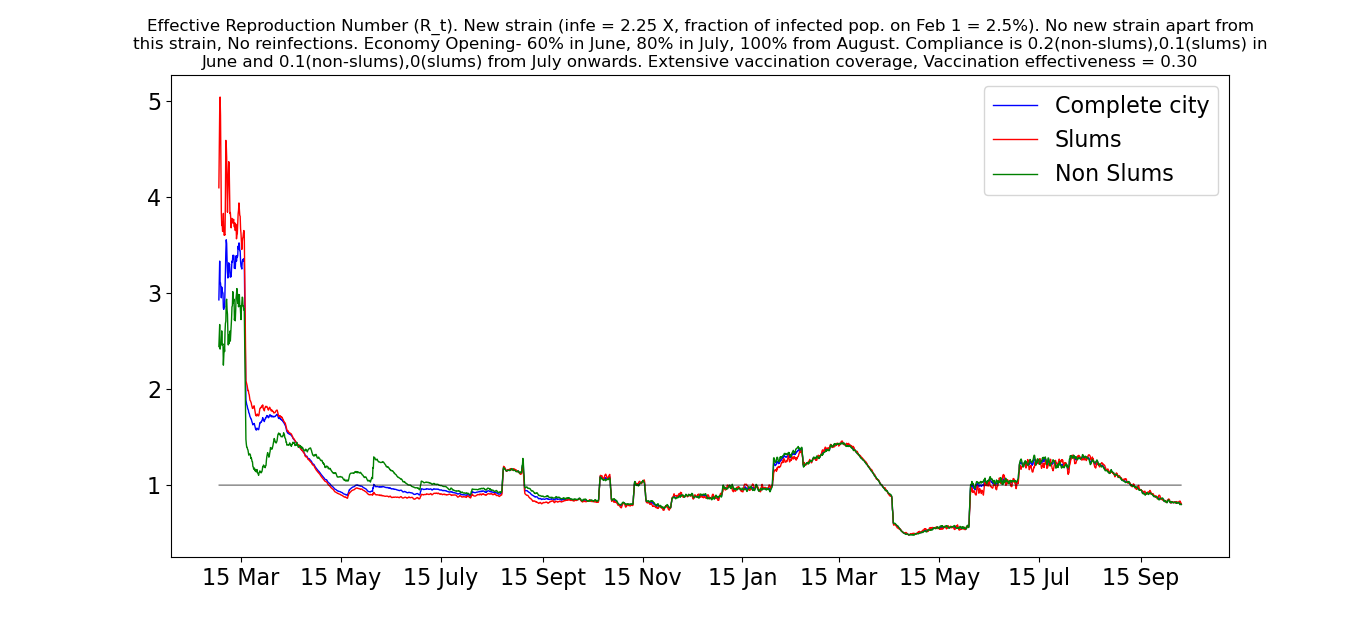}
      \caption{\small 
      Effective reproduction number ($R_t$) for complete city, slums and non-slums. New strain (infectiousness  = 2.25 X, fraction of infected pop. on Feb 1 = 2.5\%). No new strain apart from this strain, no reinfections. City mobility 
       - 60\% in June, 80\% in July, 100\% from August. Compliance is 0.2 (non-slums), 0.1 (slums) in June and 0.1 (non-slums), 0 (slums) from July onwards. Extensive vaccination coverage, Vaccination effectiveness = 0.30. 
 }\label{R_t_1}

      \centering
     \includegraphics[width=\linewidth]{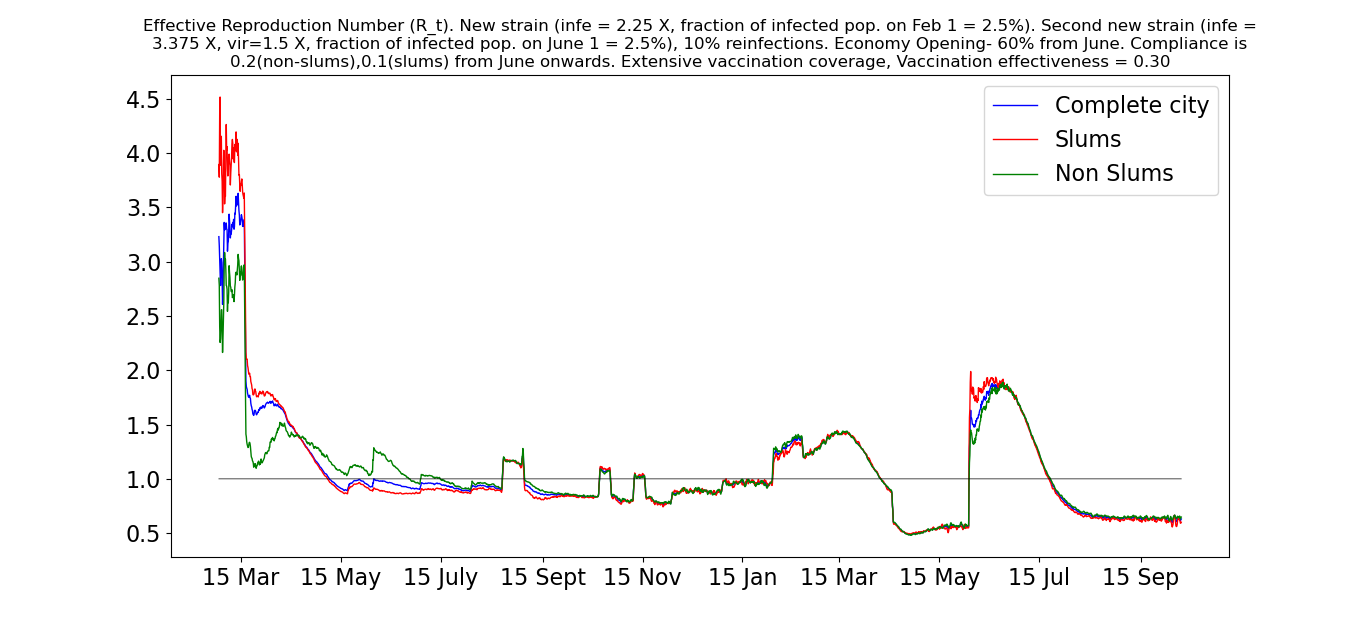}
      \caption{\small 
      Effective reproduction number ($R_t$) for complete city, slums and non-slums. New strain (infectiousness = 2.25 X, fraction of infected population on Feb 1 = 2.5\%). Second new strain (infectiousness = 3.375 X, virulence = 1.5 X, fraction of infected population on June 1 = 2.5\%), 10\% reinfections. City mobility - 60\% from June. Compliance is 0.2 (non-slums), 0.1 (slums) from June. Extensive vaccination coverage, Vaccination effectiveness = 0.30. 
 }\label{R_t_19} 
  \end{figure}

\begin{figure}
      \centering
     \includegraphics[width=\linewidth]{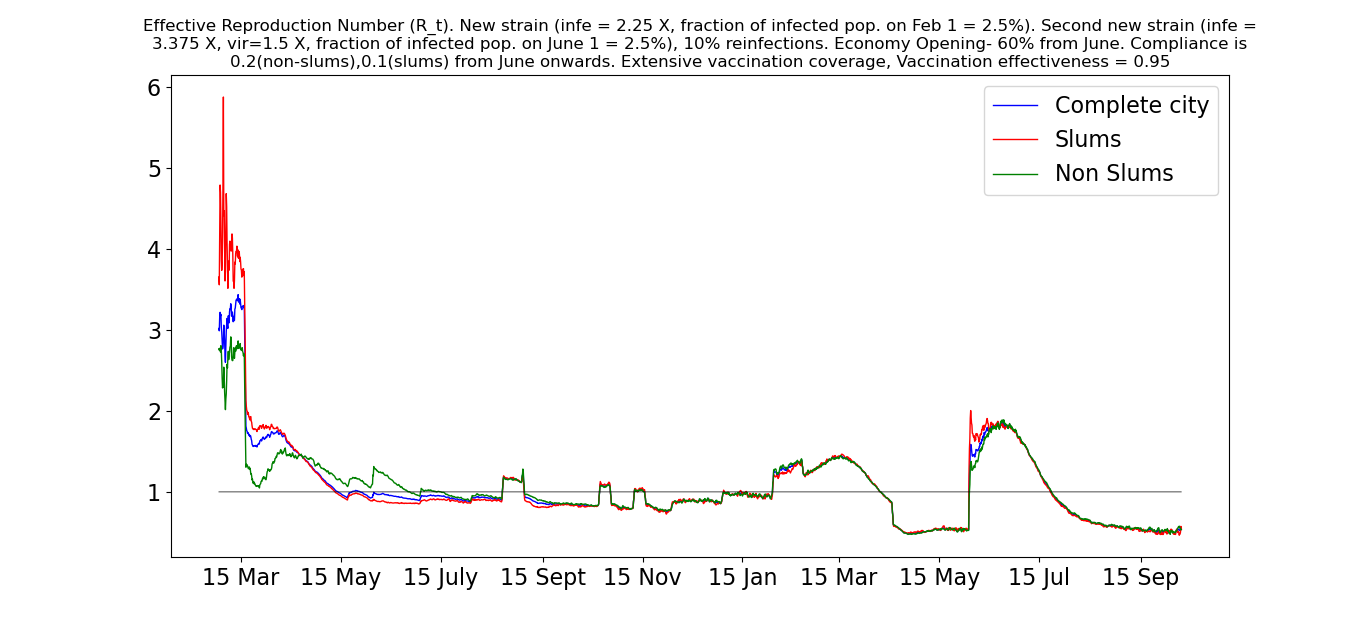}
      \caption{\small 
      Effective reproduction number ($R_t$) for complete city, slums and non-slums. New strain (infectiousness = 2.25 X, fraction of infected population on Feb 1 = 2.5\%). Second new strain (infectiousness = 3.375 X, virulence = 1.5X, fraction of infected population on June 1 = 2.5\%), 10\% reinfections. city mobility - 60\% from June. Compliance is 0.2 (non-slums), 0.1 (slums) from June. Extensive vaccination coverage, Vaccination effectiveness = 0.95.
 }\label{R_t_21} 

  \end{figure}

In Figure \ref{R_t_1},  \ref{R_t_19} and \ref{R_t_21}, we report the $R_t$ values in the three scenarios considered in Figure \ref{peak_fatalities}. Specifically, we display the following 3 scenarios:
\begin{itemize}
    \item
    In Figure \ref{R_t_all}, we considered a scenario
  where the reinfections were mild so that they affect the fatality figures negligibly and 
 where the new variants (beyond the existing delta variant) had little impact and vaccination effectiveness was 0.75. Here we consider same scenario but with low vaccination effectiveness (0.3) and observe a mild increase in $R_t$ in the coming months. Recall that in the above scenario we had assumed that the city is opened up  at 60\% level in June, 80\% in July and 100\% thereafter.
The compliance is 20\% in non-slums and 10\% in slums in June. This reduces to 10\% and zero July onwards.
    \item 
    We also consider a pessimistic scenario where reinfections are 10\% and there is a new variant (apart from delta variant) which is 3.375 times more infectious and 1.5 times more virulent than original strain. Vaccination effectiveness is assumed to be 0.3. In this scenario we have assumed that the city is opened up at 60\% level from June onwards.
The compliance is 20\% in non-slums and 10\% in slums June onwards. In this scenario the $R_t$ rises dramatically 
in June.
    \item
    The scenario considered here is same as the previous one but with a significantly higher vaccination effectiveness (0.95) resulting in a mild reduction in $R_t$ in June.
\end{itemize}

\bigskip
\section*{Acknowledgments}
We thank our colleagues Prahladh Harsha, Ramprasad Saptharishi and   Piyush Srivastava for many useful suggestions.  
We thank them as well as our IISc collaborators
R. Sundaresan,  P. Patil, N. Rathod, A. Sarath, S. Sriram, and N. Vaidhiyan for their tireless efforts
in developing the IISc-TIFR Simulation model \cite{City_Simulator_IISc_TIFR_2020}
and their key role
in our earlier reports on  Mumbai. 

We thank IDFC Institute for sponsoring Daksh Mittal's work with the TIFR COVID-19 City-Scale Simulation Team.

We thank 
World Health Organization, Regional office for South-East Asia, WHO-SEARO, team 
for our discussions and the team's feedback.

We  acknowledge the support of A.T.E. Chandra Foundation for this research.
We thank Amit Chandra for his useful feedback.

We further  acknowledge the support of the Department of Atomic Energy, Government of India, to TIFR under project no. 12-R\&D-TFR-5.01-0500.


\end{document}